\documentclass[12pt]{article}
\usepackage{graphicx}
\usepackage{amsmath}
\usepackage[utf8]{inputenc}
\usepackage[hidelinks]{hyperref}
\usepackage{float}

\newcommand{\ttbar}{t\smash{$\bar{\text{t}}$}}
\newcommand{\ttz}{\ttbar Z}
\newcommand{\sqrts}{$\sqrt{s}=13$\,TeV}
\newcommand{\perfb}{\,fb\textsuperscript{-1}}
\newcommand{\perab}{\,ab\textsuperscript{-1}}
\newcommand{\ptz}{$p_T(\text{Z})$}
\newcommand{\cts}{$\cos\theta^\star_{\text{Z}}$}
\newcommand{\ie}{\textit{i.e.}}
\newcommand{\tz}{top quark--Z~boson}

\begin{document}
\begin{titlepage}
\rightline{\begin{tabular}{l}
    CMS-CR-2019/256 \\ 
    November 28, 2019 
\end{tabular}}

\vfill
\begin{center}\Large 
    Differential measurement of \ttz{} production \\
    at \sqrts{} with the CMS experiment
\end{center}
\vfill
\begin{center}
    \href{mailto:joscha.knolle@cern.ch}{\textsc{Joscha Knolle}} \\
    \textit{Deutsches Elektronen-Synchrotron, Hamburg, Germany}
\end{center}
\begin{center}
    \textsc{on behalf of the CMS Collaboration}
\end{center}
\vfill
\begin{quotation} 
Measurements of top quark pair production in association with a Z~boson (\ttz) provide direct tests of the coupling of top quarks and Z~bosons, as well as the description of the production mechanism in QCD. The measurements can be used to constrain possible contributions from anomalous \tz{} couplings. A first differential \ttz{} production measurement has been performed by the CMS experiment using events with three leptons collected in 2016 \& 2017. The differential cross sections are reported as functions of the Z~boson transverse momentum, and of the angle between the Z~boson and one of its decay leptons, boosted to the Z~boson's rest frame. The results are compared to state-of-the-art theory predictions.
\end{quotation}
\vfill
\begin{quotation}\begin{center}
    PRESENTED AT
\end{center}\bigskip\begin{center}\large
    $12^\mathrm{th}$ International Workshop on Top Quark Physics\\
    Beijing, China, September 22--27, 2019
\end{center}\end{quotation}
\vfill
\end{titlepage}
\def\thefootnote{\fnsymbol{footnote}}
\setcounter{footnote}{0}

\section{Introduction}

The associated production of a top quark pair with a Z~boson (\ttz) in proton-proton collisions at \sqrts{} has been measured by the CMS experiment~\cite{CMS-Experiment} at the CERN LHC using 77.5\perfb{} of data collected in 2016 \& 2017. The publication comprises an inclusive cross section measurement, LHC's first differential cross section measurement of \ttz{} production, and an interpretation in the framework of the standard model (SM) effective field theory~\cite{TOP-18-009}. This contribution presents details and results of the differential \ttz{} production cross section measurement.

\section{Theory predictions}

\begin{figure}[h!]
\centering
\includegraphics[width=0.4\textwidth]{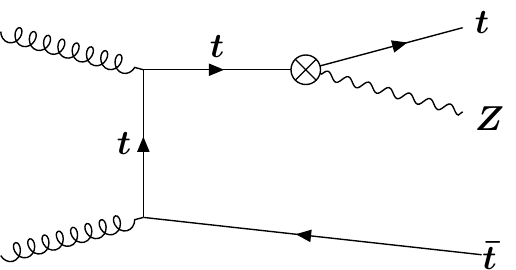}
\caption{Feynman diagram for \ttz{} production at leading order via strong and electroweak interactions. The \tz{} vertex is highlighted.}
\label{fig:feynman}
\end{figure}

The \ttz{} production process is sensitive to the electroweak \tz{} coupling, illustrated by the example Feynman diagram in Fig.~\ref{fig:feynman}. Its measurement provides the first direct probe of this coupling.

New physics beyond the SM can lead to sizeable modifications of the Lorentz structure of the \tz{} coupling. As discussed in Ref.~\cite{Roentsch15}, a differential measurement of the kinematic properties of the Z~boson allows to put constraints such anomalous couplings.

Additionally, the differential measurement can be used to probe the available description of the QCD production mechanism. In the SM, a description of \ttz{} production at next-to-leading order (NLO) accuracy in QCD and electroweak theory is available~\cite{Frederix18}. These calculations have been extended with soft gluon corrections at next-to-next-to-leading logarithmic (NNLL) accuracy~\cite{Kulesza18}.

\section{Event selection}

The differential measurement makes use of the three-lepton final state, where the Z~boson and one top quark decay leptonically (`lepton' always refers to electrons and muons), and the other top quark decays hadronically.

Events are recorded with one-, two-, and three-lepton triggers, and selected with three high-$p_T$ leptons that pass the tight working point of a dedicated identification MVA trained to separate prompt leptons (from direct decays of Z~and W~bosons) from non-prompt leptons (misidentified jets, and leptons from heavy hadron decays).

The leptonically decaying Z~boson is reconstructed from an opposite-sign same-flavour lepton pair with an invariant mass $m(\ell^+\ell^-)$ compatible with the Z~boson mass, $|m(\ell^+\ell^-)-m_{\text{Z}}|<10$\,GeV. Events without a Z~boson candidate are rejected.

To suppress background processes without a \ttbar{} pair, at least three high-$p_T$ jets are required, of which at least one is required to pass a b-tagging criterion.

Background processes without three prompt leptons (mainly \ttbar{} and Z+jets production) are estimated from control regions in data, while all other background processes (mainly tZq, tWZ, and WZ production) are estimated from Monte Carlo (MC) simulations.

\section{Unfolding procedure}

Two distributions are measured at the level of the reconstructed objects: the transverse momentum of the reconstructed Z~boson, \ptz, and the angle between the negatively charged lepton from the Z~boson decay and the reconstructed Z~boson, boosted to the Z~boson's rest frame, \cts. The events are categorized in four bins in both observables. For \ptz, the last bin has no upper limit, but for the purpose of display and normalization an upper limit is used.

To correct for effects from detector acceptance and efficiency, object resolutions, event selection and reconstruction, an unfolding procedure is applied. MC simulated \ttz{} events are used to derive the response of the measurement, \ie{} the quantitative relation between the observable at parton-level (after all QCD and electroweak radiation) and at reconstructed level. The same binning is used for the parton-level observables.

A very good resolution of the two observables is found due to good lepton resolutions of the CMS detector. The purity, \ie{} the fraction of events in a reconstructed bin that are also in the same bin at parton-level, and the stability, \ie{} the fraction of events in a parton-level bin that are reconstructed in the same bin, are larger than 94\,\% in all bins of both observables. This results in a response matrix with very small off-diagonal entries.

Inversion of the response matrix, without a regularization to suppress statistical fluctuations, is sufficient to perform the unfolding procedure. It is implemented using the \texttt{TUnfold} software~\cite{TUnfold}.

\section{Results}

\begin{figure}[H]
\centering
\includegraphics[width=0.44\textwidth]{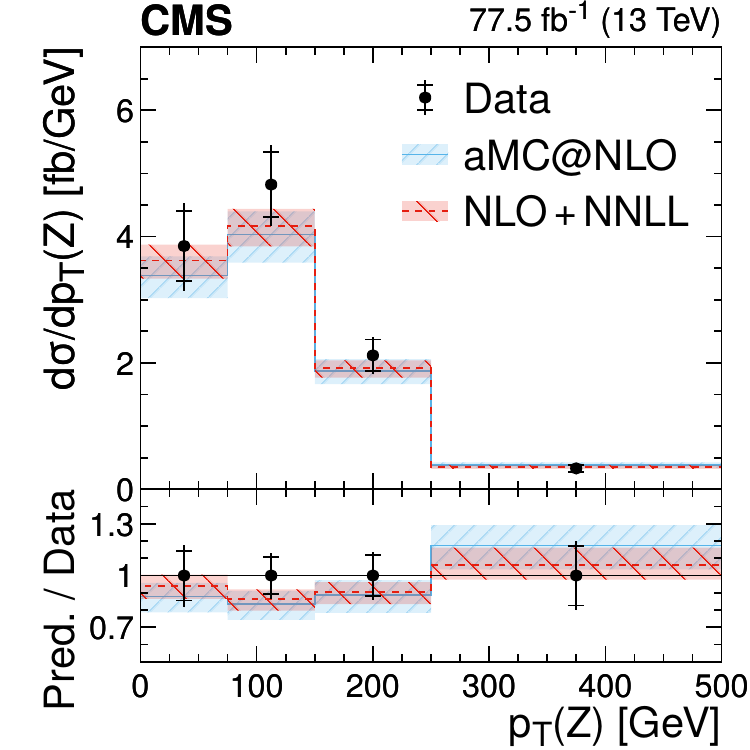}
\hspace{0.05\textwidth}
\includegraphics[width=0.44\textwidth]{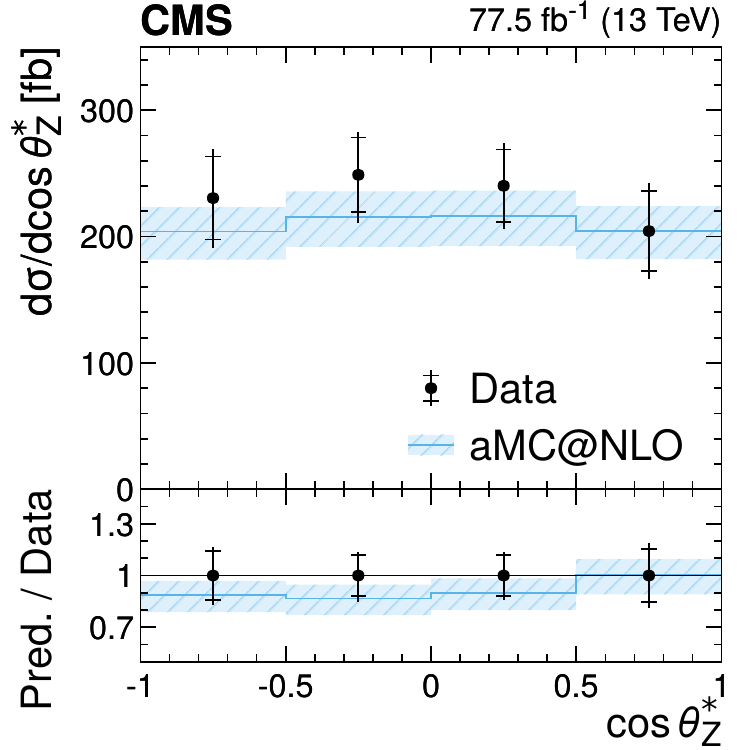} \\[0.05\textwidth]
\includegraphics[width=0.44\textwidth]{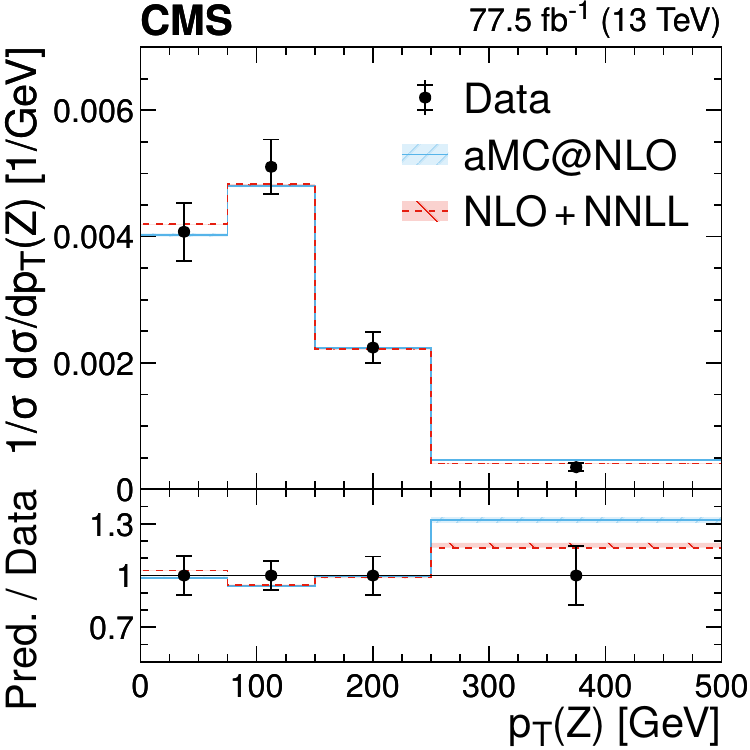}
\hspace{0.05\textwidth}
\includegraphics[width=0.44\textwidth]{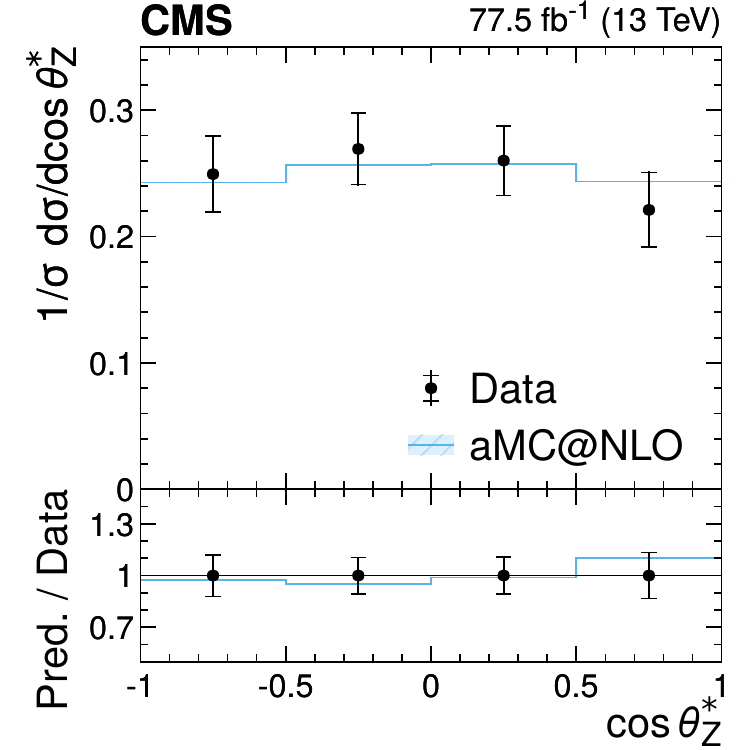}
\caption{Absolute (upper row) and normalized (lower row) measured cross sections as functions of \ptz{} (left) and \cts{} (right) \cite{TOP-18-009}. The inner (outer) error bars indicate the statistical (total) uncertainty of the measurement. The solid blue (dashed red) line indicates the SM prediction at NLO accuracy~\cite{Frederix18} (NLO+NNLL accuracy~\cite{Kulesza19}).}
\label{fig:results}
\end{figure}

Figure~\ref{fig:results} shows the measured differential cross sections for both observables, after scaling the unfolded number of events with the Z~boson branching ratio and the luminosity. The statistical uncertainty dominates the uncertainty in each bin of both observables. A comparison with SM predictions for both distributions shows good agreement with the measured cross sections.

\section{Outlook}

The CMS experiment has performed a first differential measurement of the \ttz{} production cross section in two leptonic observables, using 77.5\perfb{} of data collected in 2016 \& 2017. Since the results are limited by the statistical uncertainty, more precise results can be expected with larger datasets. Ref.~\cite{FTR-18-036} discusses the prospects of this measurement with 3\perab{} of data to be collected at the high-luminosity LHC.

Future studies will exploit a kinematic reconstruction of the \ttbar{} system. This will allow to measure distributions of observables involving the top quarks.

\bigskip\bigskip
\begin{center}\large\bfseries
    ACKNOWLEDGEMENTS
\end{center}
I am grateful to the PIER Helmholtz Graduate School for their financial support to my participation at this conference.

\end{document}